# Drying of Nanoporous Media by Concurrent Drainage and Evaporation: A Pore Network Modeling Study

Haiyi Wu,[1] Chao Fang,[1] Rui Wu,[2] and Rui Qiao[1,*]

[1]Department of Mechanical Engineering, Virginia Tech, Blacksburg, VA 24061, USA

[2]School of Mechanical Engineering, Shanghai Jiao Tong University, Shanghai, 200240, China

**Abstract.** Drainage and evaporation can occur simultaneously during the drying of porous media, but the interactions between these processes and their effects on drying are rarely studied. In this work, we develop a pore network model that considers drainage, evaporation, and rarefied multi-component gas transport in nanopores. Using this model, we investigate the drying of a liquid solvent-saturated nanoporous medium enabled by the flow of purge gas through it. Simulations show that drying progresses in three stages, and the solvent removal by drainage effects (evaporation effects) becomes increasingly weak (strong) as drying progresses through these stages. Interestingly, drainage can contribute considerably to solvent removal even after evaporation effects become very strong, especially when the applied pressure difference across the porous medium is low. We show that these phenomena are the results of the coupling between the drainage and evaporation effects and this coupling depends on the operating conditions and the stage of drying.

**Keywords**: porous media; drying; pore network models; multiphase flows; multicomponent gas transport

[*] To whom correspondence should be addressed. Email: ruiqiao@vt.edu.

# I. Introduction

Drying of porous media including building materials and pharmaceuticals is important in engineering applications ranging from material manufacturing and oil extraction to soil remediation.[1-3, 4] Although many technologies including thermal drying and infrared radiation drying have been developed,[5] they often suffer from limitations such as high energy cost, low throughput, *etc*. Hence, there is a long-standing need to improve existing methods or to develop new technologies to overcome these limitations. Addressing this need through trial-and-error experiments is often costly and ineffective, and numerical modeling can be helpful. Modeling of the drying of porous media, however, is challenging because coupled multiphase heat and mass transfer at scales from nanometer to centimeters must be simulated.[6-7]

To tackle the multi-faceted (multiphase, multiscale, and multiphysics) drying processes in porous media, many numerical models have been developed. These models can be loosely classified into two categories: macrohomogeneous models and pore scale models. In macrohomogeneous models, the porous media are treated as a continuum with volume-averaged or homogenized properties. These models can deal with large scale problems easily.[8-9] However, because the microstructures of porous media and the physical processes in them (e.g., viscous fingering and corner flows) are not resolved but modeled heuristically using empirical or experimental sub-models, they offer limited insight into the fundamental physics of the drying process and often lack predictive power in novel situations. In pore scale models, the heat and mass transfer in the microscale geometry of porous media are considered. In some models, the original heat and mass transfer equations such as the Navier-Stokes equation are solved in porous media.[10,11,12] In other models,[13] porous media are represented using various building blocks and the transport processes are modeled based on rules derived from fundamental heat and mass transfer laws. Amount these models, pore network models have received much attention because they offer a good balance between capturing pore scale physics (hence drying) and being computationally efficient.

In pore network models, the pore space is simplified as discrete elements consisting of pore bodies interconnected with each other by pore throats. The first pore network model for studying drying was built in 1954[13] and many improvements have been made since then.[14-19] Key to these models is the consideration of fluid and mass transfer at the pore scale and in the interconnected pore systems. For example, the transport of vapor obeys convection-diffusion equations and the continuity equation is applied to the liquid phase and non-condensable gas phase;[14, 17] the pattern of receding liquid-vapor interfaces due to evaporation has been described using the invasion percolation (IP) model.[20-21] While existing pore network models have been successful in analyzing many porous media drying problems, new problems with unique features not explored previously continue to appear and the existing models need to be extended to study these problems.

In this work, we extend pore network models to study the drying of nanoporous media assisted by purge gas flowing through them.

Figure 1 show a schematic of the drying problem studied here, which is part of the solvent recovery step in the dewatering-by-displacement technology.[22] A bed of particles with diameter often smaller than 2 μm is initially saturated with a volatile solvent (e.g., pentane) and a purge gas is driven through the particle bed to remove the solvent. At the beginning, the injected purge gas penetrates into the particle bed to displace the liquid solvent, just as in the conventional drainage process. Eventually, the purge gas breaks through the particle bed and gas transport pathways are formed across the porous matrix. After this, evaporation occurs within the porous matrix and the vaporized solvent is "flushed" downstream by the purge gas. Meanwhile, the purge gas continues to drive the remaining liquids out of the porous matrix. Although the extensive pore network modeling in the literature has led to useful fundamental understanding of porous media drying, they offer limited insights for this drying problem because of its several unique features:

1. Drying is assisted by the flow of purge gas *through* the porous media, which differs from the widely studied scenario where drying is assisted by gas streams flowing *over* the porous media's surface;
2. The drainage of liquid solvents and evaporation of solvents can occur concurrently in the porous media and become tightly coupled;
3. Because of the small particle size, the pores between them can have diameter of a few hundred nanometers.[23] Therefore, the gas slippage and Knudsen effects, rarely considered in previous drying research, can no longer be neglected in the gas transport process;
4. The liquid solvent is highly volatile and their vapor pressure can be comparable to that of the purge gas. The multicomponent gas transport behavior can involve strong couplings between the purge gas and vaporized solvents, which cannot be described well using the classical convection-diffusion equations.

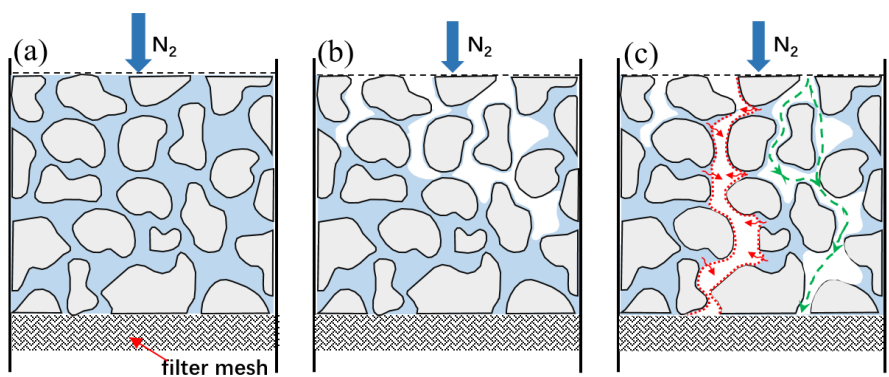

**Figure 1**. *Drying of a nanoporous medium with purge gas flowing through it*. **(a)** The porous medium, a particle bed, is initially saturated with a volatile liquid solvent. **(b)** The purge gas displaces liquid solvents from the porous medium, and the solvent evaporation is negligible. **(c)** The purge gas breaks through the porous medium and evaporation of the liquid solvents becomes important; drainage and evaporation can occur concurrently and become strongly coupled.

In this work, we study the drying of nanoporous media assisted by purge gas flowing *through* them using a pore network model. The rest of the manuscript is structured as follows: In Section II, we detail our model that accounts for the concurrent the drainage and evaporation processes during drying and the other unique features mentioned above. In Section III, we present the simulation results for drying, reporting the evolution of the drying rate and discussing the couplings between the drainage and evaporation effects. Finally, conclusions are presented in Section IV.

## II. Model Formulation

To study the drying problem shown in Fig. 1, we assume that the nanoporous media is initially ($t = 0$) fully saturated with a volatile liquid solvent. The solvent liquids have a zero contact angle on solid surfaces and the purge gas has negligible solubility in the solvent liquids. Below, we first present the setup of the pore network, then discuss the physical and numerical models, and finally present the numerical algorithm.

### A. Pore network

Since we focus on the coupling between drainage and evaporation, which is the most unique aspect of the dyring problem shown in Fig. 1, the geometry of the porous media is simplified as a two-dimensional (2D) rectangular block represented by a pore network.[24-25] The pore network lattice consists of pore bodies connected to their four neighbors via pore throats (see Fig. 2a). The upstream (downstream) boundary of the pore network is connected with a upstream (downstream) reservoir. Periodic boundary conditions are applied on the left and right boundaries. The pore bodies have cubic shape and the pore throats have circular cross-sections. A schematic of two pore bodies connected via a pore throat is presented in Fig. 2b. Following previous work,[24-25] the size distribution of pore bodies and pore throats both follow a truncated log-normal distribution:

$$f(r_i) = \frac{\sqrt{2}\exp(-0.5(\frac{\ln\frac{r_i}{r_m}}{\sigma}))}{\sqrt{\pi\sigma^2}\cdot r_i(\text{erf}(\ln\frac{r_{max}}{r_m}/\sqrt{2\sigma^2})-\text{erf}(\ln\frac{r_{min}}{r_m}/\sqrt{2\sigma^2}))} \quad (1)$$

where $r_i$ is the radius of the inscribed sphere of a cubic pore body $i$ (see Fig. 2b); in other words, the length of the pore body is $2r_i$. $\sigma$ is the standard deviation. $r_m, r_{min}$ and $r_{max}$ are the mean of the inscribed sphere radius, the lower bound and upper bound of the truncation, respectively. The spacing between adjacent layers in the pore network in the *x*- and *y*-directions are denoted as $\Delta S_{x,i}$ and $\Delta S_{y,i}$, respectively:

$$\Delta S_{x,i} = a_x \max\{r(i,j) + r(i+1,j), \text{ for } j = 1:n_y\}, \quad i = 1:n_x \quad (2a)$$
$$\Delta S_{y,j} = a_y \max\{r(i,j) + r(i,j+1), \text{ for } i = 1:n_x\}, \quad j = 1:n_y \quad (2b)$$

where $\alpha_x$ and $\alpha_y$ are prefactors set to $a_x = a_y = 1.2$ and $n_x$ ($n_y$) is the number of pore bodies in the *x*- (*y*-) direction. The length of the pore throats is then determined via

$$\begin{cases} l_{x,i} = \Delta S_{x,i} - (r(i,j) + r(i+1,j)) \text{ for } j = 1:n_y, & i = 1:n_x \\ l_{y,j} = \Delta S_{x,j} - (r(i,j) + r(i,j+1)) \text{ for } i = 1:n_x, & j = 1:n_y \end{cases} \quad \begin{matrix} (3a) \\ (3b) \end{matrix}$$

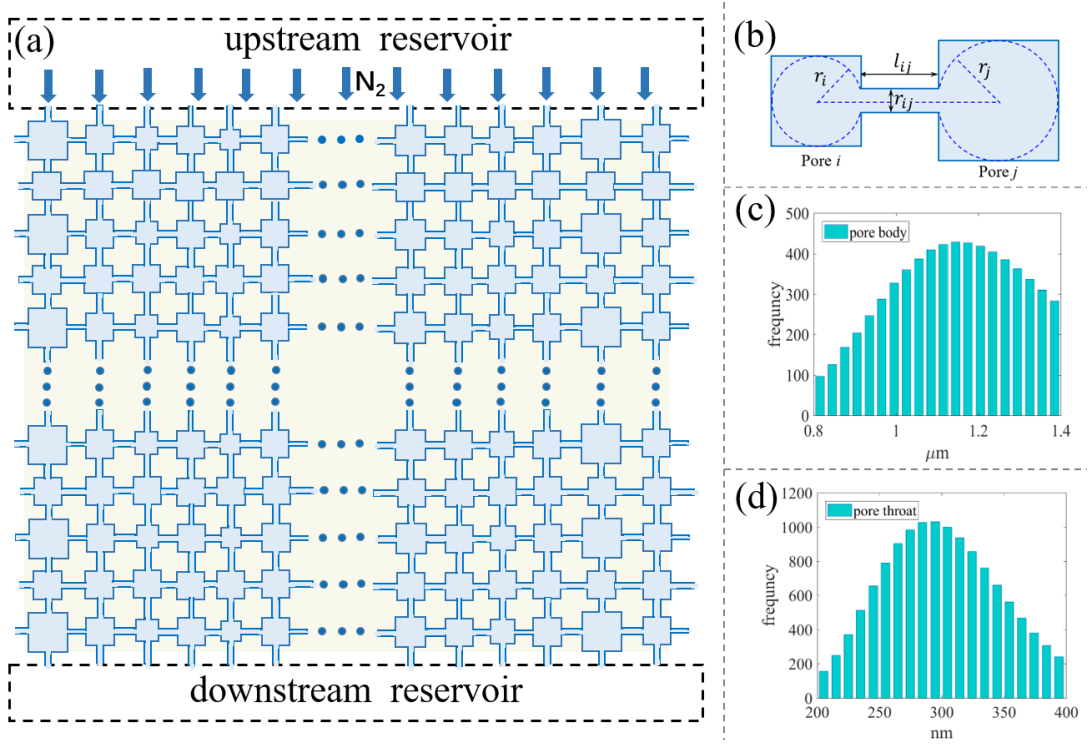

**Figure 2**. The pore network model. **(a)** A schematic of the pore network. **(b)** A sketch of two pore bodies connected by a pore throat. **(c)** The pore body size distribution. **(d)** The pore throat size distribution.

## B. Physical and numerical models

Below, we present the models for the drying of nanoporous media assisted by purge gas flowing through them. Key assumptions for model development, the models for liquid and vapor transport, the models for liquid vaporization, initial and boundary conditions, rules for events in the pore network model, and the computational algorithms are discussed.

### B.1 Assumptions

The following assumptions are made in our pore network model:

1. Pore bodies have finite volume but zero hydraulic resistance for liquid and gas transport.
2. Pore throats have negligible volume and their hydraulic resistance corresponds to fully developed flows in ducts with the same size. A pore throat has only two states: filled with liquid or filled with gas. Meniscus is not tracked inside pore throats and the time required for filling/emptying a single pore throat by liquid/gas is negligible.
3. Liquid solvents are incompressible and the solid phase is considered as a rigid body.

4. For the gas and liquid flows in pore throats, the inertia effects are negligible.
5. Gravity effects are negligible because the drying process occurs at microscale.
6. Drying occurs under isothermal conditions.
7. The vaporization of liquid solvents at the corner of a pore body can occur if more than two throats connected with that pore body is occupied by the gas phase. For pore bodies in which vaporization does not occur, the density of the solvent vapor is set to its saturation density at the same temperature.

These assumptions have been adopted extensively in prior pore network simulations of porous media drying because they allow capture the essential physics of the drying process to be modeled with modest cost and computer memory.[17, 21, 24, 26-27]

### B.2 Transport model

We adopt the two-pressure algorithm[24, 28-29] for solving the pressure field for both liquid and gas phases. Hereafter, variables corresponding to the liquid and gas phases are identified using superscripts "*l*" and "*g*", respectively. Because the gas phase is generally a two-component mixture of the purge gas (species 1) and the solvent vapor (species 2), variables corresponding these components are identified using superscripts "*g1*" and "*g2*", respectively. The local capillary pressures in the pore bodies are defined and approximated as[24, 29]

$$P_i^c(S_i^l) = P_i^g - P_i^l = \frac{2\gamma}{r_i(1-\exp(-6.83 S_i^l))} \quad (4)$$

where $P_i^g$, $P_i^l$, $P_i^c$ and $S_i^l$ are the gas phase pressure, liquid phase pressure, capillary pressure, and the saturation of the liquid phase in the pore body $i$. $\gamma$ is the interfacial tension.

*Volume balance:* A volumetric flux is assigned to a pore throat *ij* for both gas and liquid phases and a volume balance is required for each pore body *i*:

$$\sum_{j=1}^{N_i}(Q_{ij}^l + Q_{ij}^g) = 0 \quad (5)$$

where $N_i$ is the number of pore bodies connected to the pore $i$, $Q_{ij}^l$ ($Q_{ij}^g$) is the volumetric flux of liquid (gas) through the pore throats *ij*.

*Liquid transport*: The volumetric flux of the liquid phase through a pore throat *ij* is

$$Q_{ij}^l = -K_{ij}^l(P_i^l - P_j^l) \quad (6)$$

where $P_i^l$ is the liquid pressure of pore body $i$ and, $K_{ij}^l$ is the conductivity of the pore throats filled with liquid. $K_{ij}^l = \frac{\pi r_{ij}^4}{8\mu_l l_{ij}}$ since we assume a Hagen-Poiseuille flow in pore throats. $\mu_l, l_{ij}, r_{ij}$ are the liquid viscosity, pore throat length, and pore throat radius, respectively.

*Gas transport:* The transport of gas through pore throats is complicated: because the partial pressure of the solvent vapor is comparable to that of the purge gas, a multicomponent transport model is required. Here, we adopt the dusty gas model (DGM), which have been widely used for multicomponent mass transfer in nanopores. The transport of an ideal mixture of $d$ gas species through a pore with a radius of $r_{ij}$ can be described using[30-31]

$$\sum_{\substack{n=1 \\ n \neq m}}^{d} \frac{x_n J_m - x_m J_n}{\rho_t D_{mn}} + \frac{J_m}{\rho_t D_{mk}} = -\frac{1}{p}\frac{dp_m}{dz} - \frac{x_m k_p}{\mu_{mix} D_{mk}}\frac{dp}{dz} \quad (m = 1, 2, \dots d) \tag{7}$$

where $J_m$, $\rho_m$, and $x_m$ are the molar flux, averaged molar density, and molar ratio of specie $m$, respectively. $p$ is the pressure of the $d$-component gas mixture and $p_m$ is the partial pressure of a specie $m$. $\rho_t$ is the total molar density, $k_p$ is the effective permeability, $\mu_{mix}$ is the mixture viscosity, $D_{mn}$ is the mutual diffusivity between species $m$ and $n$. $D_{mk} = \frac{2r_{ij}}{3}\sqrt{\frac{8R_gT}{\pi M_m}}$ is the Knudsen diffusivity of species $m$ ($M_m$ is the molar mass, $R_g$ is the ideal gas constant and $T$ is the temperature). When applied to a binary mixture of purge gas (species 1) and solvent vapor (species 2) that behaves ideally, the molar flux of each gas species can be reorganized and simplified as

$$J_1 = \rho_1 u_1 = -\left(\left(\frac{(x_1 D_{2k} D_{1k} + D_{12} D_{1k})/R_gT}{x_2 D_{1k} + x_1 D_{2k} + D_{12}} + \frac{k_p \rho_1}{\mu_{mix}}\right)\frac{dp_1}{dz} + \left(\frac{x_1 D_{2k} D_{1k}/R_gT}{x_2 D_{1k} + x_1 D_{2k} + D_{12}} + \frac{k_p \rho_1}{\mu_{mix}}\right)\frac{dp_2}{dz}\right) \tag{8a}$$

$$J_2 = \rho_2 u_2 = -\left(\left(\frac{x_2 D_{2k} D_{1k}/R_gT}{x_2 D_{1k} + x_1 D_{2k} + D_{12}} + \frac{k_p \rho_2}{\mu_{mix}}\right)\frac{dp_1}{dz} + \left(\frac{(x_2 D_{2k} D_{1k} + D_{12} D_{2k})/R_gT}{x_2 D_{1k} + x_1 D_{2k} + D_{12}} + \frac{k_p \rho_2}{\mu_{mix}}\right)\frac{dp_2}{dz}\right) \tag{8b}$$

The overall gas volume flux $Q_{ij}^g$ through pore throat $ij$ can be wrote as

$$Q_{ij}^g = \frac{\pi r_{ij}^2 (J_{1,ij} + J_{2,ij})}{\rho_{ij}^g} = -\left(K_{ij}^A\left(P_i^g - P_j^g\right) + K_{ij}^B\left(P_i^{g2} - P_j^{g2}\right)\right) \tag{9}$$

where $K_{ij}^A = \left(\frac{(D_{2k,ij}D_{1k,ij} + D_{12,ij}D_{2k,ij})/R_gT}{x_{2,ij}D_{1k,ij} + x_{1,ij}D_{2k,ij} + D_{12,ij}} + \frac{k_p \rho_{ij}^g}{\mu_{mix}}\right)\frac{\pi r_{ij}^2}{\rho_{ij}^g l_{ij}}$ and $K_{ij}^B = \left(\frac{D_{12,ij}(D_{2k,ij} - D_{1k,ij})/R_gT}{x_{2,ij}D_{1k,ij} + x_{1,ij}D_{2k,ij} + D_{12,ij}}\right)\frac{\pi r_{ij}^2}{\rho_{ij}^g l_{ij}}$ are transport coefficients. $P_i^{g2}$ is the vapor's partial pressure in pore body $i$, $\rho_{ij}^g = 0.5(\rho_i^g + \rho_j^g)$ is the average molar density in pores $i$ and $j$. Substituting Eqs. (6) and (9) into Eq. (5), we obtain

$$\sum_{j=1}^{N_i}\left(K_{ij}^l\left(P_i^l - P_j^l\right) + K_{ij}^g\left(P_i^g - P_j^g\right) + K_{ij}^{g2}\left(P_i^{g2} - P_j^{g2}\right)\right) = 0 \tag{10}$$

Eq. (10) can be reformulated using the average pressure of the pore body $\tilde{P}_i = (1 - S_i^l)P_i^g + S_i^l P_i^l$ as

$$\sum_{j=1}^{N_i}\left(K_{ij}^l + K_{ij}^g\right)\left(\tilde{P}_i - \tilde{P}_j\right) = -\sum_{j=1}^{N_i}\left(K_i^c P_i^c - K_j^c P_j^c + K_{ij}^{g2}\left(P_i^{g2} - P_j^{g2}\right)\right) \tag{11}$$

where $K_i^c = K_{ij}^g S_i^l - K_{ij}^l(1 - S_i^l)$ and $K_j^c = K_{ij}^g S_j^l - K_{ij}^l(1 - S_j^l)$.

Using the kinetic theory of gas transport inside nanopores and considering the first order slippage effect, the effective permeability for gas through a pore throat $ij$ is given by[32]

$$k_p = \frac{r_{ij}^2}{8}\left(1 + 4\text{Kn}\left(\frac{2}{\sigma_v} - 1\right)\right) \quad (12)$$

where $\sigma_v$ is the tangential momentum accommodation coefficient. We assume $\sigma_v = 1$, which corresponds to a rough pore surface that reflects all molecules diffusively. $\text{Kn} = \lambda/2r_{ij}$ is the Knudsen number of the gas mixture ($\lambda$ is the mean free path for the gas mixture, see below), which defines the ratio of the Knudsen diffusion to self-diffusion in a bulk gas.

*Properties of gas mixture*: To complete the above gas transport model, the mutual diffusivity, mixture viscosity, and the mean free path of the gas mixture are needed. The mutual diffusivity in a pore throat $ij$ is approximated using the empirical formula[30, 33]

$$D_{12} = \left(D_{x_1 \to 1}\right)^{x_1}\left(D_{x_1 \to 0}\right)^{(1-x_1)} \quad (13)$$

where $x_1 = 0.5(x_{1,i} + x_{1,j})$ is the molar fraction of species 1 in a pore throat $ij$, which is taken as the average of the molar faction of species 1 in pore body $i$ ($x_{1,i}$) and pore body $j$ ($x_{1,j}$). The bracketed terms are the infinite dilution values for the Maxwell-Stefan diffusivity at either end of the composition range and are given by $D_{x_i \to 1} = \frac{2}{3}\frac{\sqrt{m_i k_B T/\pi}}{M_i \sigma_{ai}\rho_i}$ ($i = 1,2$), where $M_i$ and $\sigma_{ai}$ are the molar mass and effective collision cross-section area of specie $i$, respectively. We approximate the viscosity of the binary gas mixture using an empirical correlation:[34]

$$\mu_{mix} = \frac{\mu_1}{1+\frac{x_2 1.385\mu_1}{x_1 \, D_{12}\rho_1}} + \frac{\mu_2}{1+\frac{x_1 1.385\mu_2}{x_2 \, D_{12}\rho_2}} \quad (14)$$

where $\mu_i$ is the viscosity for a pure species $i$. The mean free path of a binary gas mixture is given by

$$\lambda = \frac{x_1 m_1 + x_2 m_2}{\sqrt{2}\sigma_{a1}^{x_1}\sigma_{a2}^{x_2}(M_1\rho_1 + M_2\rho_2)} \quad (15)$$

where $m_i$ ($i = 1,2$) are molecular mass for the species $i$.

**B.3 Liquid vaporization model**

The evaporation of liquid solvents at the corner of pore bodies and the vapor's subsequent transport to the pore center are essential steps of the vaporization process induced by purge gas. Since we assume that drying proceeds under the isothermal condition, the vaporization rate is limited by the vapor transport from the liquid surface toward the pore's center rather than the kinetics of liquid evaporation. Therefore, the vapor density on the liquid surface is equal to the liquid's saturation vapor density $\rho_{sat}$ and the vaporization from the liquid surface inside a pore body $i$ is given as:

$$\dot{m}_i^{evp} = \beta A_s (\rho_{sat} - \rho_i^{g2}) \tag{16}$$

where $\beta$ is the mass transfer coefficient. $A_s$ is the interfacial area between the liquid solvent and the gas phase inside the pore body and it is given by:[24]

$$A_s = \begin{cases} \frac{144}{\pi} r_i^2 (1-S_i^l)^{\frac{2}{3}}, & S_i^l \geq 0.476 \\ \frac{12\pi r_i \gamma}{P_c} - \frac{8\pi\gamma^2}{P_c^2}, & S_i^l < 0.476 \end{cases} \tag{17}$$

The mass transfer of vapor from the surface of liquids at pore corners to the pore interior is characterized using the Sherwood number $Sh = 2\beta r_i/D_s$, which depends on the pore's shape. Here $Sh$ is taken as 2.98, which corresponds to square-shaped pores.[35]

With the above vaporization model, the mass balance of the solvent vapor in a pore body $i$ is given by

$$V_i^g \Delta \rho_i^{g2} / \Delta t = \sum_{j=1}^{N_i} \pi r_{ij}^2 J_{ij}^{g2} + A_s \beta (\rho_{sat} - \rho_i^{g2}) \tag{18a}$$

$$J_{ij}^{g2} = -\left( C_{ij}^g (P_i^g - P_j^g) + C_{ij}^{g2} (\rho_i^{g2} - \rho_j^{g2}) \right) \tag{18b}$$

where $V_i^g = V_i(1 - S_i^l)$ is the volume of the gas phase inside a pore body $i$ and $\Delta t$ is the time step (see B.5 for details), $C_{ij}^g = \frac{x_{2,ij} D_{1k,ij} D_{2k,ij}/R_g T}{x_{1,ij} D_{2k,ij} + x_{2,ij} D_{1k,ij} + D_{12,ij}} + \frac{k_p}{\mu_{mix}} \rho_{ij}^{g2}$ and $C_{ij}^{g2} = \frac{D_{2k,ij} D_{12,ij}}{x_{1,ij} D_{2k,ij} + x_{2,ij} D_{1k,ij} + D_{12,ij}}$ are transport coefficients.

### B.4 Initial and boundary conditions

Initially, the pore network is fully saturated with the liquid solvents:

$$S_i^l(t=0) = 1.0 \tag{19}$$

During drying, the pressure at the pore network's inlet ($y=0$ in Fig. 2) is fixed to that of the upstream purge gas reservoir $P_u$ and no solvent vapor is transported into the pore from the upstream reservoir:

$$y = 0: P^g|_{y=0} = P_u \tag{20a}$$

$$y = 0^-: J^{g2} = 0 \tag{20b}$$

The pressure at the pore network's outlet ($y = L$ in Fig. 2) is fixed to that at the downstream reservoir ($P_d$):

$$y = L: P^g|_{y=L} = P_d \tag{20c}$$

$$y = L: \frac{\partial \rho^{g2}}{\partial y} = 0 \tag{20d}$$

**B.5 Rules**

1. *Threshold pressure for pore throats.* Because the contact angle of solvent liquids on the solid surfaces is zero, the threshold pressure for pore throats filled with liquids is

$$P_{th,ij} = \frac{2\gamma}{r_{ij}} \tag{21}$$

A pore throat filled with liquid can be invaded by the gas phase if the capillary pressure in the neighboring pore body exceeds the threshold pressure of this pore throat: [24]

$$P_c(S_i^l) > P_{th,ij} \text{ or } P_c(S_j^l) > P_{th,ij} \tag{22}$$

Because the liquid saturation $S_i^l$ in a pore body can be reduced by vaporization of the liquid solvents in it, Eq. (22) implies that evaporation can indirectly affect the invasion of a pore throat by liquid solvents or gas.

2. *Re-imbibition of pore throats.* If a pore body $i$ is refilled with liquid solvent ($S_i^l = 1.0$), an empty throat (i.e., a throat free of liquid) $ij$ can be re-imbibed with the liquid under the condition[29]

$$P_i^g > P_j^g \tag{23}$$

where $P_i^g$ is the gas pressure of the target pore body and $P_j^g$ is the gas pressure in the pore body connected to the target pore body by the throat $ij$.

3. *Selection of time step.* The time step for each iteration in the pore network simulation is determined based on the local time steps required for draining a pore body to the critical state, i.e., when the local capillary pressure satisfies $P_c(S_i^l) = \min\{P_{th,ij}\}$. Since the imbibition process is allowed to occur locally in pore bodies, the local time step can be determined by the time required to refill the pore body[24]:

$$\Delta t_i = \begin{cases} \frac{V_i}{\sum_{j=1}^{N_i} Q_{ij}^l}(S_i^l - S_{i,ct}) & \text{local drainage, } \sum_{j=1}^{N_i} Q_{ij}^l < 0 \\ \frac{V_i}{\sum_{j=1}^{N_i} Q_{ij}^l}(1 - S_i^l) & \text{local imbibition, } \sum_{j=1}^{N_i} Q_{ij}^l > 0 \end{cases} \tag{24}$$

where $S_{i,ct}$ is the saturation corresponding to the critical state. Using the $P_c - S_i^l$ relation by Eq.(4), the critical saturation can be given as

$$S_{i,ct} = -\frac{1}{6.83}\ln(1 - \frac{2\gamma}{r_i \min\{\Delta P_{ud}, \min\{P_{th,ij}\}\}}) \tag{25}$$

where $\Delta P_{ud} = P_u - P_d$ is the global pressure difference between the upstream and downstream reservoir. $\min\{P_{th,ij}\}$ is the minimum threshold pressure among the liquid throats connected to the target pore body. Then the global time step is selected as the minimum of all local time steps, i.e., $\Delta t = \min\{\Delta t_i\}$.

## C. Computational algorithms

Starting from the initial conditions given by Eq. (19), Eqs. (4) and (11) are combined with the boundary conditions given by Eq. (20) to solve the pressure field in the pore network using the direct sparse-matrix solver (DSS).[36] Using this pressure field, the solvent vapor density for the next time step is computed by solving Eqs. (17) and (18) subject to the boundary conditions in Eq. (20) using an implicit scheme. These calculations are repeated till the entire pore network is free of liquid solvents. These and other details of the computational algorithms for solving the pore network model are summarized in Fig. 3.

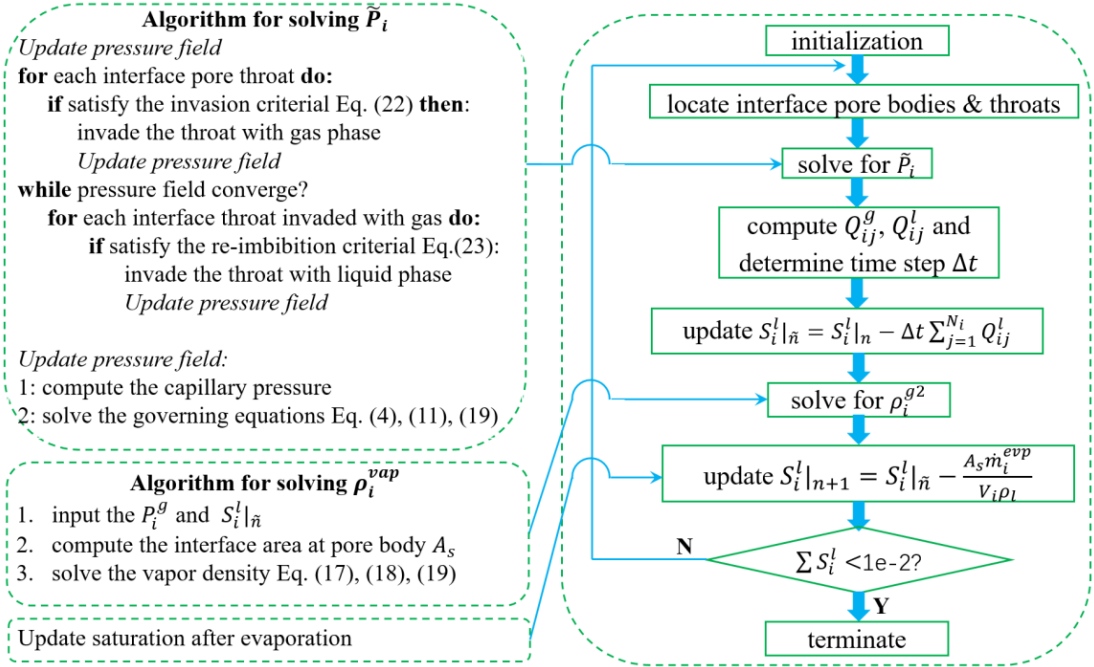

**Figure 3**. The computational procedure and algorithm for solving the pore network model in which drainage and evaporation can occur concurrently.

## III. Results and Discussion

We consider the drying of a porous coal cake initially saturated with liquid solvent pentane by the flow of a purge gas $N_2$ through the cake. Because coal particles' diameter is often less than a few micrometers, the size of the pores in the coal cake can be a few hundred nanometers. Here, the coal cake is represented using a pore network consisting of 80 × 80 pore bodies, with each pore body connected to its 4 nearest pore bodies as shown in Fig. 2a. The size of pore bodies follows the truncated log-normal distribution with $\sigma_b = 0.22$, $r_{b,m} = 1.2$ μm, $r_{b,min} = 0.8$ μm and $r_{b,max} = 1.4$ μm. The size of pore throats follows the same distribution with $\sigma_t = 0.18$, $r_{t,m} = 300$ nm, $r_{t,min} = 200$ nm and $r_{t,max} = 400$ nm. The size distributions of pore bodies and throats are shown in Fig. 2c and 2d. The pore bodies and throats generated above are randomly packed into the pore network.

The system temperature is 300 K. The material properties of the purge gas ($N_2$) and solvent (pentane) at this temperature are summarized in Table 1. Note that the effective cross-section areas of the $N_2$ and pentane molecules ($\sigma_{a,1}$ and $\sigma_{a,2}$) are calculated based on the results in Ref. 37.

Table 1. Properties of the purge gas $N_2$ (species 1) and solvent (species 2).

| | |
|---|---|
| $m_1 = 4.65 \times 10^{-26}$ kg | $\sigma_{a,1} = 0.43$ nm$^2$ |
| $m_2 = 12.0 \times 10^{-26}$ kg | $\sigma_{a,2} = 2.09$ nm$^2$ |
| $\rho_{sat,2} = 28.1$ mol/m$^3$ | $\mu_1 = 17.8$ µPa·s (gas state) |
| $P_{sat,2} = 0.7$ bar | $\mu_2 = 70.0$ µPa·s (gas state) |
| $\rho_2 = 626.0$ kg/m$^3$ (liquid state) | $\mu_1 = 224.0$ µPa·s (liquid state) |
| $\gamma = 15$ mN/m | |

The pressure in the reservoir downstream the pore network, $P_d$, is 1.0 bar. For selection of the upstream gas pressure ($P_u$), we identify a minimal pressure difference as

$$\Delta P_{ud}^{min} = \frac{2\gamma}{r_{t,min}} \quad (28)$$

When $P_u - P_d = \Delta P_{ud}^{min}$, not all all pores inside the pore network can be emptied by drainage because the pressure drop along individual throats is smaller than the global pressure difference across the entire pore network. Nevertheless, $\Delta P_{ud}^{min}$ provides an indication of the ability of the applied pressure difference to drain liquids from the pore network. For the pore network studied here, $\Delta P_{ud}^{min} = 1.5$ bar. In our simulations, we select $\Delta P_{ud} = P_u - P_d = 1.4, 1.5,$ and $2.0$ bar to study how the applied pressure difference affects the drying behavior. These values are within the range used in the dewatering-by-displacement technology.[22]

## A. Macroscopic drying behaviors

To quantify how a porous medium initially saturated with liquid solvents is dried, we define a *degree of drying* $\delta$ using $\delta(t) = 1 - \bar{S}(t)$, where $\bar{S}$ is the ratio of the liquid solvent mass at a time $t$ and the initial solvent mass in the porous medium. In purge gas-assisted drying of a porous medium, solvent can be removed as liquid through drainage (termed drainage effect) and by the transport of vaporized solvents out of the porous medium (termed evaporation effect). We thus decompose the net solvent removal rate $\dot{F}_{net}$ from the porous medium as $\dot{F}_{net} = \dot{F}_{drn} + \dot{F}_{evp}$, where $\dot{F}_{drn}$ and $\dot{F}_{evp}$ are solvent fluxes at the porous medium' outlet due to the drainage and evaporation effects, respectively. To quantify the relative importance of the drainage and evaporation effects, we further define an evaporation-to-drainage ratio as $\alpha = \dot{F}_{evp}/\dot{F}_{drn}$.

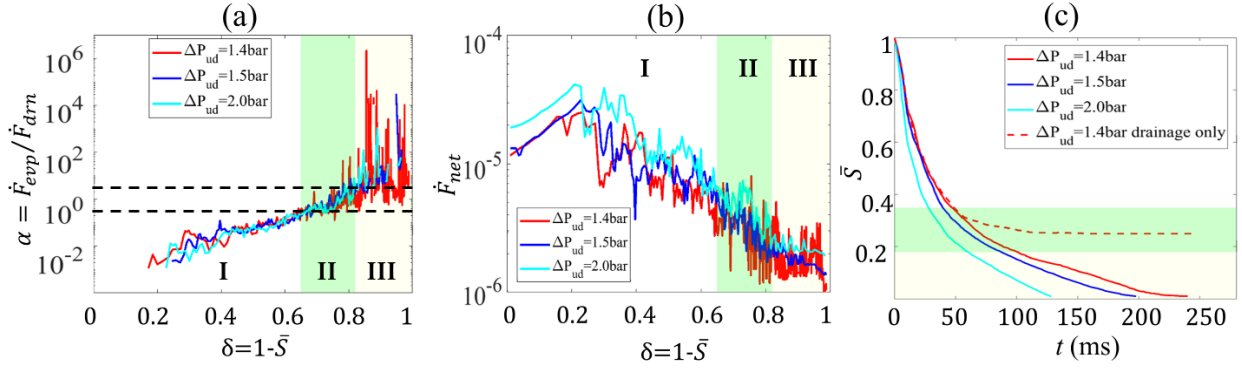

**Figure 4**. *The overall drying behavior at different applied pressure $\Delta P_{ud}$.* (**a-b**) The evolution of the ratio of removal rates of solvent as vapor and liquid (a) and the total solvent removal rate (b) as a function of the degree of drying δ. (**c**) The evolution of the average liquid saturation in the porous medium. The dashed line corresponds to the simulation in which evaporation is disabled.

Figure 4a shows the evolution of $\alpha$ during drying under three pressure differences $\Delta P_{ud}$. As drying proceeds, $\alpha$ generally increases, i.e., the evaporation effect becomes more and more important. Base on the relative importance of the drainage and evaporation effects, the drying process under all $\Delta P_{ud}$ can be divided into three stages: *stage I* with $\alpha < 0.3$, during which the solvent removal is dominated by the drainage effect; *stage II* with $0.3 < \alpha < 3.0$, during which the drainage and evaporation effects are comparable; *stage III* with $\alpha$ mostly larger than 3, during which the drying is largely dominated by the evaporation effects. For all $\Delta P_{ud}$, the degree of drying spanned by each stage is similar, i.e., $\delta \lesssim 0.65$ in stage I, $0.65 \lesssim \delta \lesssim 0.82$ in stage II, and $\delta \gtrsim 0.82$ in stage III. During stage III, $\alpha$ fluctuates notably and can reach ~0.3-1 (i.e., drainage notably contributes to solvent removal), and this is especially pronounced for the smallest $\Delta P_{ud}$ studied (1.4 bar). The contribution of drainage to the solvent removal during each stage is summarized in Table 2. We observe that, in the first two stages, the contribution of drainage to solvent removal is similar for all $\Delta P_{ud}$. However, in stage III, drainage contributes more significantly at lower $\Delta P_{ud}$, e.g., the contribution of drainage to solvent removal increases from 6.2% at $\Delta P_{ud} = 2.0$ bar to 21.4% at $\Delta P_{ud} = 1.4$ bar. Nevertheless, because most of the solvents are removed during stage I, the overall contribution of drainage to solvent removal does not differ greatly for the $\Delta P_{ud}$ studied: as $\Delta P_{ud}$ decreases from 2.0 to 1.4 bar, the contribution of drainage increases only by 1.5%. Nevertheless, because solvent removal by drainage avoids the energy cost due to the heat of vaporization needed in evaporation-induced solvent removal, even a small increase of drainage's contribution is desirable, especially in large-scale applications.

Figure 4b shows the evolution of the net solvent removal rate $\dot{F}_{net}$ during drying. Overall, $\dot{F}_{net}$ is higher for larger $\Delta P_{ud}$. The $\dot{F}_{net}$ curves at different $\Delta P_{ud}$ exhibit similarities. First, $\dot{F}_{net}$ fluctuates notably except at the beginning of stage I. Second, the time evolution of $\dot{F}_{net}$ is similar: in stage I, $\dot{F}_{net}$ increases first and then decreases. $\dot{F}_{net}$ continues to decrease in stage II but plateaus in stage III. Significant fluctuation of $\dot{F}_{net}$ is observed in the stage III, especially at $\Delta P_{ud} = 1.4$ bar.

**Table 2**. The contribution of drainage to the solvent removal and the duration of each drying stage.

| | | $\Delta P_{ud} = 1.4\ bar$ | $\Delta P_{ud} = 1.5\ bar$ | $\Delta P_{ud} = 2.0\ bar$ |
|---|---|---|---|---|
| contribution of drainage to the solvent removal | stage I | 96.1% | 96.6% | 96.9% |
| | stage II | 57.8% | 53.1% | 61.0% |
| | stage III | 21.4% | 8.9% | 6.2% |
| | total | 75.9% | 73.3% | 74.4% |
| duration (ms) | stage I | 46.82 | 39.58 | 27.77 |
| | stage II | 53.91 | 47.77 | 32.22 |
| | stage III | 139.30 | 110.50 | 68.46 |
| | total | 240.03 | 197.85 | 128.45 |

Figure 4c shows the time evolution of the averaged solvent liquid saturation in the porous medium. The liquid saturation decreases sharply during stage I, but the decrease of saturation slows down in stage II and even more so in stage III. These observations are consistent with the fact that, as drying proceeds, drying becomes controlled more by the evaporation effects and the solvent removal rate decreases (see Fig. 4a and 4b). Note that the evaporation effects are essential for the complete removal of solvents. To appreciate this, we perform simulations in which the vaporization of liquid solvents is disabled and only drainage is allowed. Comparison of the result of this simulation (the dashed line in Fig. 4c) with the above result shows that, as the average liquid saturation reduces to ~0.3, the solvent removal starts to be affected by the evaporation effect. In absence of the evaporation effects, a significant fraction of the liquid solvents remains trapped in the porous medium.

Figure 4c shows that the evaporation effects-dominated stage III lasts longer compared to the other two stages, i.e., the throughput of drying is limited by the removal of solvents by evaporation effects. This is seen more clearly in Table 2, in which the time corresponding to each drying stage is listed. The time needed for the stage III is especially long under the applied pressure difference of 1.4 bar, which is partly caused by the smaller driving force for gas transport through the pore network.

## B. Microscopic drying processes

The existence of three drying stages, the contributions and interplay of drainage and evaporation effects during drying, and finally the evolution of the drying rate revealed in Fig. 4 can be understood by examining the microscopic processes in the pore network during drying. Figure 5 shows the snapshots of the liquid saturation distribution in the pore network at four representative degrees of drying ($\delta$) when the applied pressure difference is $\Delta P_{ud} = 1.5$ bar. Similar results are observed for other $\Delta P_{ud}$ and are not shown here.

Once the pressure difference is imposed across the pore network, the purge gas starts to invade into the pore network. The purge gas travels preferentially through pathways with wider pore throats and displaces the liquids in a piecemeal manner. This is evident in Fig. 5a ($\delta$ =0.16), where finger-shaped gas paths are observed. The formation of these paths, often referred to as viscous fingering,[38-39] is generally considered as the onset and evolution of instabilities when a more viscous fluid (here, liquid solvent) is displaced by a less viscous fluid (here, purge gas). As the viscous fingers grow, the network formed by liquid-saturated pore bodies is fragmented into liquid clusters (hereafter, a liquid cluster is defined as a collection of pore bodies fully saturated by liquids and connected continuously by liquid-filled throats). In particular, the initial main liquid cluster spanning the entire pore network is fragmented into small liquid clusters, which are broken into even smaller clusters later. Many small liquid clusters, along with some large liquid clusters, appear inside the system (see Fig. 5b, where $\delta$ =0.5). During the early part of this process, as viscous fingers move toward but have not yet reached the pore network's outlet, the resistance for drainage decreases and the drainage rate increases, which is consistent with the initial increase of the drying rate (see Fig. 4b, $\delta \lesssim 0.2 - 0.3$). After the purge gas breaks through the pore network, drainage from the pore network's outlet becomes more limited. Therefore, the drying rate decreases as shown in Fig. 4b and the evaporation effects begin to contribute to solvent removal. However, until drainage pathways are diminished by the fingering of gas pathways, the removal of solvents is dominated by drainage and drying is in the stage I identified in Fig. 4a.

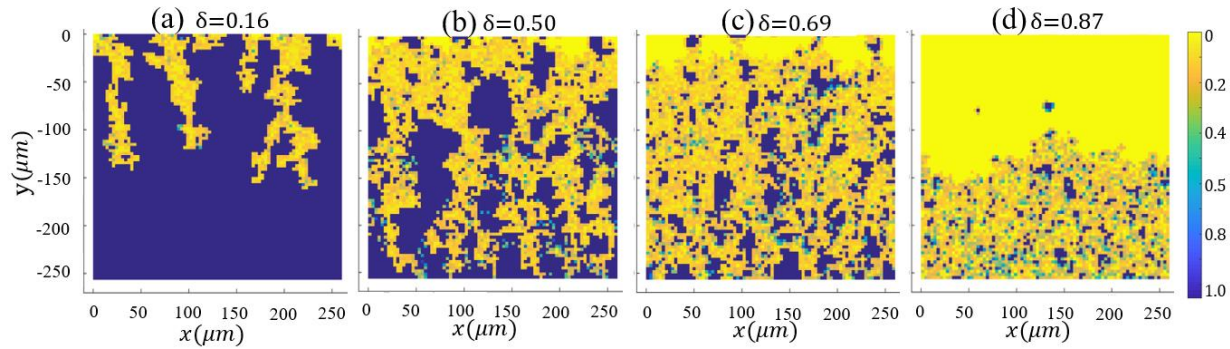

**Figure 5**. The evolution of liquid saturation in the pore network when the applied pressure difference $\Delta P_{ud}$ is 1.5 bar.

As drying proceeds, the liquid saturation throughout the pore network reduces, and this is evident in the snapshot in Fig. 5c ($\delta$=0.69). Consequently, the drainage of liquid solvents through the network is reduced while more pathways for vaporized solvents to be "flushed" out of the network emerge. The solvent removal by the evaporation effects thus increases. The contributions of drainage and evaporation effects eventually becomes comparable and drying enters stage II. Because drainage is much more effective in removing solvent (note that the density of liquid solvents is >100 times larger than the density of vaporized solvents), the net drying rate continues to decrease in stage II.

In stage II of the drying process, a drying front, behind which few liquid clusters exist, appears near the pore network's inlet (see Fig. 5c) and moves downstream as drying proceeds. However, it important to note that solvents are removed from individual pores both near and *far ahead of* this drying front. As drying continues, more and more pores near the network's outlet become only partially occupied by liquid solvents. Therefore, solvent removal by drainage (evaporation effects) diminishes (becomes significant), and drying enters stage III as shown in Fig. 4a.

In stage III, the drying rate is quite stable for $\Delta P_{ud}$ = 1.5 and 2.0 bar (see Fig. 4b). Even for $\Delta P_{ud}$ = 1.4 bar, the drying rate largely fluctuates around a constant value. These trends can be understood as follows. At this stage, small liquid clusters are mostly trapped in pores with narrow throats. The pathway for gas (purge gas + solvent vapor) transport evolves only slowly and the gas flow rate is relatively stable. Because the gas phase at the pore network's exit is generally saturated with the solvent vapor (except toward the very end of drying), the net flux of vaporized solvents out of the porous medium is rather stable. Because drying is dominated by the vaporized solvents, the total drying rate is relatively stable.

## C. The interplay between drainage and evaporation effects

The above analysis highlights the role of drainage and evaporation effects in purge gas-assisted drying of porous media and helps understand the main features of the macroscopic drying behavior. However, some features of the drying data shown in Fig. 4 still cannot be readily understood. For example, a notable drainage flux can still appear after the evaporation effects start to dominate drying (see Fig. 4a and Table 2), which is also manifested as the spikes in the net drying rate in stage III (see Fig. 4b). These features essentially arise from the coupling between the drainage and evaporation effects.

To understand the coupling between drainage and evaporation effects and its impact on the macroscopic drying behavior, it is instructive to examine the evolution of liquid clusters in the pore network, which are affected oppositely by the two processes: in drainage, purge gas breaks through liquid clusters to fragment them into small clusters; the evaporation in pore bodies tends to eliminate small liquid clusters. Figure 6a shows the evolution of the number of liquid clusters $N_c$ in the pore network. In stage I, evaporation effects are minor and $N_c$ increases rapidly as viscous fingers break through increasingly smaller clusters. In stage II, fragmentation of liquid clusters by purge gas "fingers" becomes weaker while the elimination of liquid clusters by evaporation effects becomes significant. The competition of these processes leads to a plateau of $N_c$. In stage III, where evaporation effects dominate, $N_c$ decreases.

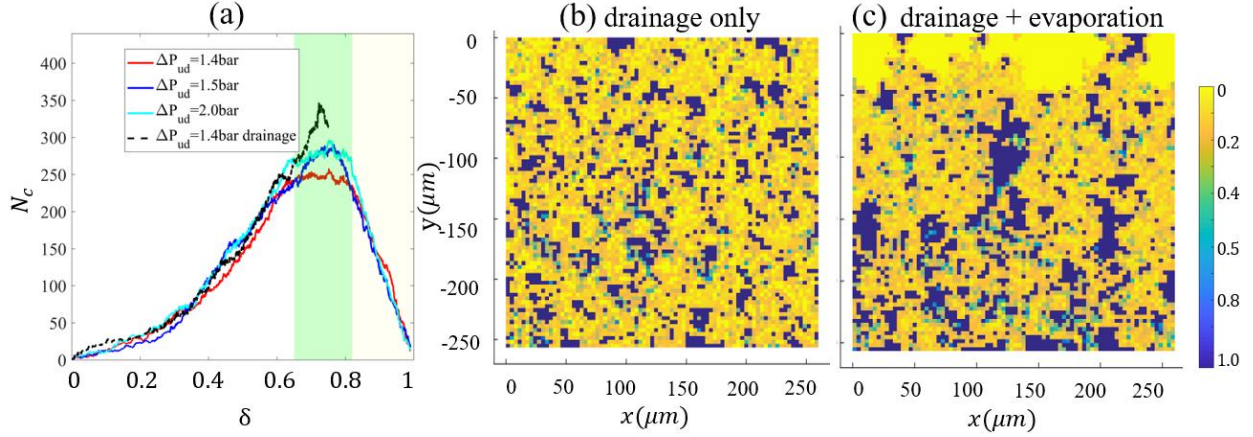

**Figure 6**. (**a**) The evolution of the number of liquid clusters in the pore network during drying and drainage-only operations. (**b-c**) The distribution of the liquid saturation in a pore network at a degree of drying of $\delta = 0.7$ as obtained from drying simulations (b) and drainage-only simulations (c). The applied pressure difference $\Delta P_{ud}$ is 1.4 bar.

To more directly appreciate the *coupling* between drainage and evaporation effects, we performed a new simulation in which the evaporation effects are turned off so that only drainage exists. The setup of the pore network is identical to the drying simulation reported above and the pressure difference $\Delta P_{ud}$ is set to 1.4 bar. Figure 6a shows that, in absence of evaporation effects, the evolution of $N_c$ is similar in the drainage and drying simulations till the degree of drying $\delta$ reaches 0.6, which is expected because evaporation effects are weak in the stage I drying. However, in absence of evaporation effects, $N_c$ increases with $\delta$ till it reaches ~0.7. After that, solvents are permanently trapped in small liquid clusters (see Fig. 6b) and $N_c$ remains constant. Importantly, in this case, numerous liquid clusters with small size are scattered throughout the pore network (see Fig. 6b), in sharp contrast with the fact that, at the same $\delta$, the liquid clusters in the simulation with both drainage and evaporation effects have much larger size and are distributed closer to the pore network's outlet (see Fig. 6c). This difference can be understood as follows. Once evaporation enables the emptying of a pore throat connected to a liquid cluster, drainage can displace the liquid in the cluster to neighboring pore bodies. This effectively drives the liquid cluster downstream and even allows it to merge with other liquid clusters and form the large liquid clusters seen in Fig. 6c.

The coupling between drainage and evaporation effects revealed through the comparison between Fig. 6b and 6c affects the macroscopic drying behavior. For example, the liquid clusters near the pore network's outlet can be removed through drainage, which helps explain why the solvent flux at the porous medium's outlet frequently includes a significant faction of liquid even in stage III of drying (see Fig. 4a, especially when $\Delta P_{ud}$ is 1.4 bar).

The extent of the coupling between drainage and evaporation effects during a drying process is affected by both the operating conditions (e.g., the applied pressure difference) and the stage of drying. To gauge the extent of this coupling, we note that, the coupling between the drainage and evaporation effects can be considered as three steps looped together. First, evaporation triggers the emptying of pore throats (see Eq. (22)) and prompts liquid to drain from pores. Second, drainage generates new gas transport pathways and affects the liquid saturation distribution in the pore network. Third, the altered gas transport pathways and liquid saturation distribution in turn affect the evaporation of liquid solvents in the partially saturated pores. These looped steps are most easily delineated for a liquid cluster surrounded by very narrow pore throats (see *Appendix*) but they are applicable to all liquid clusters in a pore network. Because the coupling loop between drainage and evaporation effects is triggered when evaporation induces the emptying of pore throats, we can infer the extent of such coupling by counting the number of its triggering events (i.e., evaporation-induced emptying of pore throat) $N_{trig}$. A larger $N_{trig}$ corresponds to a more extensive coupling between the drainage and evaporation effects during drying.

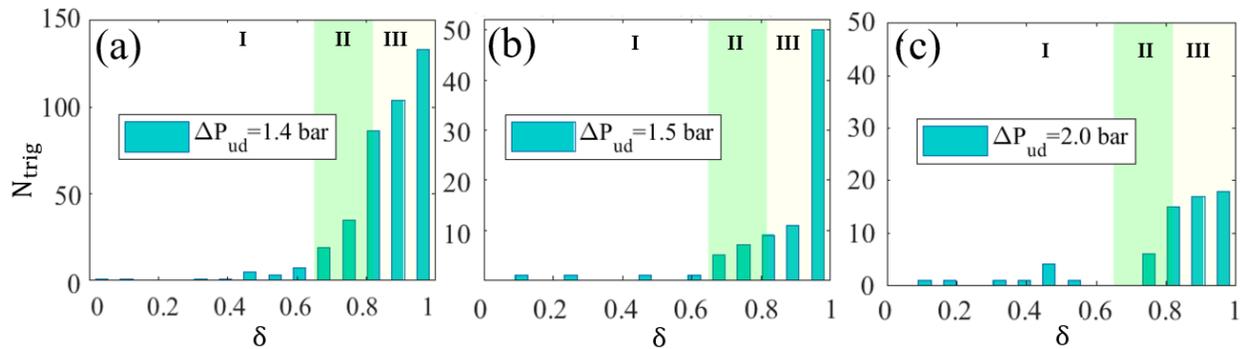

**Figure 7.** *The statistics of the triggering events (evaporation-induced emptying of pore throats) in the drainage-evaporation coupling loops during drying.* (**a-c**) The histogram of the number of triggering events as a function of the degree of drying $\delta$ under different applied pressure differences.

Figure 7a, 7b, and 7c show the histogram of $N_{\text{trig}}$ under an applied pressure difference of $\Delta P_{ud} = 1.4$, 1.5, and 2.0 bar, respectively. There are few triggering events for drainage-evaporation coupling in stage I, but these events start to appear in stage II, when the drainage and evaporation effects become comparable. In stage III, when the evaporation effect dominates, many triggering events are detected. The number of triggering events is largest at $\Delta P_{ud} = 1.4$ bar, i.e., the coupling between the drainage and evaporation effects is most extensive at this applied pressure difference. As discussed in section III.B, liquid clusters can be removed in two processes in presence of the coupling effects: (1) liquid clusters can migrate downstream, merge with other liquid clusters, and be drained out of the pore network at its outlet; (2) liquid clusters can be directly removed from the porous media via evaporation effects. Because the coupling of drainage and evaporation effects is more extensive for $\Delta P_{ud} = 1.4$ bar, the first process is strongest at this applied pressure difference. This explains why, during stage III of drying, the removal of liquid solvents by drainage from the pore network's outlet is more significant for $\Delta P_{ud} = 1.4$ bar than for $\Delta P_{ud} = 1.5$ and 2.0 bar (see Fig. 4a and Table 2).

## IV. Conclusions

In this work, we develop a new pore network model for purge gas-assisted drying of nanoporous media. The model considers multiphase and multiphysics processes such as liquid drainage, evaporation, and transport of gas mixtures through nanopores. Solutions of the model indicate that the drying process can be divided into three stages. In stage I, drying is dominated by drainage, in which gas flow displaces the liquid solvent in a fingering pattern. The net drying flux increases before the purge gas breaks through the porous medium and decreases after that. During this process the number of liquid clusters in the porous medium increases. In stage II, the evaporation effects become comparable to the drainage effect. While drainage tends to increase the number of liquid clusters in the porous medium, evaporation tends to reduce the number of liquid clusters. Hence, the number of liquid clusters in the porous medium plateaus. As drying proceeds to stage III, the liquid removal is largely dominated by the evaporation effects, although notable drainage can also occur at the porous medium's outlet, especially when the pressure driving the purge gas is low.

A key feature of the present drying process is the coupling between the drainage and evaporation effects. Evaporation can trigger three-step coupling loops that begin with the emptying of pore throats, continue with the drainage of liquids and fragmentation of liquid clusters, and continue with evaporation in the newly created partially saturated pore bodies. Because of the interplay between drainage and evaporation effects in these coupling loops, studying purge gas-assisted drying by dividing it into a drainage period and an evaporation period and studying them separately is generally inadequate (cf. the vastly different liquid saturation distributions in Fig. 6b and 6c). The coupling between drainage and evaporation effects depends on operating conditions, e.g., it is extensive in a porous medium when the applied pressure difference is comparable to the minimum threshold pressure for gas to invade the narrowest pore throats. The extent of coupling greatly affects the evolution of the liquid clusters in the porous medium and consequently the drying behavior, e.g., as the coupling becomes more extensive, liquid removal through drainage in stage III of the drying process increases, which helps increases the energy efficiency of drying. The insight on the coupling between the drainage and evaporation effects in purge gas-assisted drying helps guide future application of this method.

## Appendix

In the main text, the coupling of drainage and evaporation effects is conceptualized as three looped steps occurring in a space initially occupied by a liquid cluster: evaporation-induced pore throat emptying, drainage through the emptied throats and alteration of liquid saturation distribution, and alteration of the evaporation in that space. Here we perform simulations to illustrate these steps in a pore network where a well-defined liquid cluster exists. The statistics of the pore bodies and throats of this pore network are the same as those in the main text except that the radius of the pore throats connecting the pore bodies in a 10×20 region (termed trapped region hereafter and is delimited using a red box in Fig. A1) to the pore bodies outside of this region are selected so that their threshold pressure is larger than 1.4 bar. Simulation is then performed with an applied pressure difference of 1.4 bar across the entire pore network. The solvent liquids in the trapped region would be permanently confined in this region without evaporation effects.

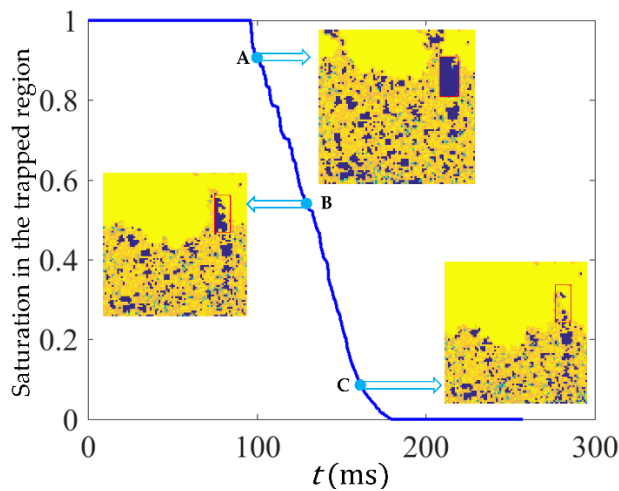

**Figure A1**. The evolution of the liquid saturation distribution inside the "trapped" region (marked using a red box), where a liquid cluster would be permanently trapped in absence of the evaporation effects.

The evolution of the liquid saturation distribution inside the trapped region is shown in Fig. A1. Due to evaporation-induced emptying of pore throats surrounding the trapped region, purge gas breaks into the liquid cluster in this region at $t \sim 95.0$ ms and displaces the liquid solvent downstream (see the snapshot at point A in Fig. A1). The gas flow then fingers through the liquid cluster in the trapped region, which in turn generates new gas pathways (see the snapshot at point B). During this process, liquids are drained out of the trapped region. Meanwhile, the liquid drainage and gas flow fingering change the distribution of the remaining liquids in and around the trapped region (see the snapshot at point C), which greatly affects the evaporation of liquids in the partially-saturated pore bodies in the trapped region. Because of the coupled drainage and evaporation effects, the liquids in the trapped region are dried out eventually.

**Acknowledgements:** The allocation of computing time by the ARC at Virginia Tech is acknowledged.

**Declarations of interest**: none.

# References

1. Panagiotou, N.; Stubos, A.; Bamopoulos, G.; Maroulis, Z., Drying Kinetics of a Multicomponent Mixture of Organic Solvents. *Drying technology* **1999**, *17*, 2107-2122.

2. Peishi, C.; Pei, D. C., A Mathematical Model of Drying Processes. *Int. J. Heat Mass Tran.* **1989**, *32*, 297-310.
3. Bruin, S.; Luyben, K. C. A. In *Drying of Food Materials: A Review of Recent Developments*, 1st. Int. Symp. Drying, Montreal, 1978, 1978.
4. Le Gallo, Y.; Le Romancer, J.; Bourbiaux, B.; Fernandes, G. In *Mass Transfer in Fractured Reservoirs During Gas Injection: Experimental and Numerical Modeling*, SPE Annual Technical Conference and Exhibition, Society of Petroleum Engineers: 1997.
5. Sharma, G.; Verma, R.; Pathare, P., Thin-Layer Infrared Radiation Drying of Onion Slices. *Journal of Food Engineering* **2005**, *67*, 361-366.
6. Mujumdar, A. S., *Handbook of Industrial Drying*; CRC press, 2014.
7. Defraeye, T., Advanced Computational Modelling for Drying Processes–a Review. *Applied Energy* **2014**, *131*, 323-344.
8. Waananen, K.; Litchfield, J.; Okos, M., Classification of Drying Models for Porous Solids. *Drying technology* **1993**, *11*, 1-40.
9. Luikov, A. V., Heat and Mass Transfer in Capillary-Porous Bodies. In *Advances in Heat Transfer*, Elsevier: 1964; Vol. 1, pp 123-184.
10. Hirt, C. W.; Nichols, B. D., Volume of Fluid (Vof) Method for the Dynamics of Free Boundaries. *J. Comput. Phys.* **1981**, *39*, 201-225.
11. Prodanović, M.; Bryant, S. L., A Level Set Method for Determining Critical Curvatures for Drainage and Imbibition. *Journal of colloid and interface science* **2006**, *304*, 442-458.
12. Aidun, C. K.; Clausen, J. R., Lattice-Boltzmann Method for Complex Flows. *Annual review of fluid mechanics* **2010**, *42*, 439-472.
13. Philip, J.; De Vries, D., Moisture Movement in Porous Materials under Temperature Gradients. *Eos, Transactions American Geophysical Union* **1957**, *38*, 222-232.
14. Yiotis, A. G.; Stubos, A. K.; Boudouvis, A. G.; Tsimpanogiannis, I. N.; Yortsos, Y. C., Pore-Network Modeling of Isothermal Drying in Porous Media. *Transport in Porous Media* **2005**, *58*, 63-86.
15. Yiotis, A. G.; Boudouvis, A. G.; Stubos, A. K.; Tsimpanogiannis, I. N.; Yortsos, Y. C., Effect of Liquid Films on the Drying of Porous Media. *Aiche Journal* **2004**, *50*, 2721-2737.
16. Prat, M., On the Influence of Pore Shape, Contact Angle and Film Flows on Drying of Capillary Porous Media. *International Journal of Heat and Mass Transfer* **2007**, *50*, 1455-1468.
17. Plourde, F.; Prat, M., Pore Network Simulations of Drying of Capillary Porous Media. Influence of Thermal Gradients. *International Journal of Heat and Mass Transfer* **2003**, *46*, 1293-1307.
18. Wu, R.; Zhao, C.; Tsotsas, E.; Kharaghani, A., Convective Drying in Thin Hydrophobic Porous Media. *Int. J. Heat Mass Tran.* **2017**, *112*, 630-642.
19. Wu, R.; Kharaghani, A.; Tsotsas, E., Two-Phase Flow with Capillary Valve Effect in Porous Media. *Chemical Engineering Science* **2016**, *139*, 241-248.
20. Wilkinson, D.; Willemsen, J. F., Invasion Percolation: A New Form of Percolation Theory. *Journal of Physics A: Mathematical and General* **1983**, *16*, 3365.
21. Lenormand, R.; Zarcone, C.; Sarr, A., Mechanisms of the Displacement of One Fluid by Another in a Network of Capillary Ducts. *J. Fluid Mech.* **1983**, *135*, 337-353.
22. Yoon, R.-H. Methods of Enhancing Fine Particle Dewatering. Patent No. 6,855,260, 2005.
23. Xiumin, J.; Chuguang, Z.; Che, Y.; Dechang, L.; Jianrong, Q.; Jubin, L., Physical Structure and Combustion Properties of Super Fine Pulverized Coal Particle. *Fuel* **2002**, *81*, 793-797.

24. Joekar-Niasar, V.; Hassanizadeh, S. M.; Dahle, H., Non-Equilibrium Effects in Capillarity and Interfacial Area in Two-Phase Flow: Dynamic Pore-Network Modelling. *J. Fluid Mech.* **2010**, *655*, 38-71.
25. Joekar-Niasar, V.; Hassanizadeh, S., Analysis of Fundamentals of Two-Phase Flow in Porous Media Using Dynamic Pore-Network Models: A Review. *Critical reviews in environmental science and technology* **2012**, *42*, 1895-1976.
26. Prat, M., Percolation Model of Drying under Isothermal Conditions in Porous Media. *International Journal of Multiphase Flow* **1993**, *19*, 691-704.
27. Gielen, T.; Hassanizadeh, S.; Leijnse, A.; Nordhaug, H., Dynamic Effects in Multiphase Flow: A Pore-Scale Network Approach. In *Upscaling Multiphase Flow in Porous Media*, Springer: 2005; pp 217-236.
28. Thompson, K. E., Pore-Scale Modeling of Fluid Transport in Disordered Fibrous Materials. *Aiche. J.* **2002**, *48*, 1369-1389.
29. Khayrat, K.; Jenny, P., A Multi-Scale Network Method for Two-Phase Flow in Porous Media. *J. Comput. Phys.* **2017**, *342*, 194-210.
30. Krishna, R.; Wesselingh, J., The Maxwell-Stefan Approach to Mass Transfer. *Chemical Engineering Science* **1997**, *52*, 861-911.
31. Mason, E. A.; Malinauskas, A., *Gas Transport in Porous Media: The Dusty-Gas Model*; Elsevier Science Ltd, 1983; Vol. 17.
32. Brown, G. P.; DiNardo, A.; Cheng, G. K.; Sherwood, T. K., The Flow of Gases in Pipes at Low Pressures. *Journal of Applied Physics* **1946**, *17*, 802-813.
33. Vignes, A., Diffusion in Binary Solutions. Variation of Diffusion Coefficient with Composition. *Industrial & Engineering Chemistry Fundamentals* **1966**, *5*, 189-199.
34. Buddenberg, J.; Wilke, C., Calculation of Gas Mixture Viscosities. *Industrial & Engineering Chemistry* **1949**, *41*, 1345-1347.
35. Young, L. C.; Finlayson, B. A., Mathematical Models of the Monolith Catalytic Converter: Part Ii. Application to Automobile Exhaust. *Aiche. J.* **1976**, *22*, 343-353.
36. Davis, T. A., *Direct Methods for Sparse Linear Systems*; Siam, 2006; Vol. 2.
37. Rothe, E. W.; Bernstein, R. B., Total Collision Cross Sections for the Interaction of Atomic Beams of Alkali Metals with Gases. *J. Chem. Phys.* **1959**, *31*, 1619-1627.
38. Liu, H.; Valocchi, A. J.; Kang, Q.; Werth, C., Pore-Scale Simulations of Gas Displacing Liquid in a Homogeneous Pore Network Using the Lattice Boltzmann Method. *Transport. Porous. Med.* **2013**, *99*, 555-580.
39. Homsy, G. M., Viscous Fingering in Porous Media. *Annual review of fluid mechanics* **1987**, *19*, 271-311.